\documentclass[useAMS,usenatbib]{mn2e} 
\usepackage{epsfig}
\usepackage{graphicx,amsmath,multicol}
\usepackage{psfrag}
\usepackage[dvipsnames]{xcolor}
\definecolor{webgreen}{rgb}{0,.5,0}
\definecolor{webbrown}{rgb}{.6,0,0}
\usepackage[pdftex,pdfpagelabels]{hyperref}
\hypersetup{%
   colorlinks=true,hyperfootnotes=false,%
   breaklinks=true,%
   plainpages=false, bookmarksnumbered, bookmarksopen=true,
bookmarksopenlevel=1,%
   urlcolor=webbrown, linkcolor=RoyalBlue, citecolor=webgreen,%
   pdftitle={PUT TITLE HERE},%
   pdfauthor={\textcopyright\ authors},%
   pdfsubject={},%
   pdfkeywords={},%
   pdfcreator={pdfLaTeX},%
   pdfproducer={LaTeX with hyperref}%
}

\newcommand{\comments}[1]{}
\newcommand       \be           {\begin{equation}}
\newcommand       \ee           {\end{equation}}
\newcommand       \ba           {\begin{eqnarray}}
\newcommand       \ea           {\end{eqnarray}}

\newcommand       \apj          {ApJ}

\newcommand       \aap          {A\&A}
\newcommand       \nat          {Nature}
\newcommand       \mnras        {MNRAS}
\newcommand       \araa         {Ann. Rev. Astr. Astr.}

\def\msun{\rm \ M_\odot}

\def\lesssim{\mathrel{\hbox{\rlap{\hbox{\lower4pt\hbox{$\sim$}}}\hbox{$<$}}}}
\def\gtrsim{\mathrel{\hbox{\rlap{\hbox{\lower4pt\hbox{$\sim$}}}\hbox{$>$}}}}

\defcitealias{mcc11}{Paper~I}

\title[Black Hole Transients induced by Cooling]
{Radiatively Inefficient Accretion Flow Simulations with Cooling: Implications for Black Hole Transients}   
\author[U.\ Das, P.\ Sharma]
{Upasana Das\thanks{upasana@physics.iisc.ernet.in} \& Prateek Sharma\thanks{prateek@physics.iisc.ernet.in} \\ Department of Physics \& Joint Astronomy Programme, Indian Institute of Science, Bangalore, India 560012}

\begin{document}

\pagerange{\pageref{firstpage}--\pageref{lastpage}} \pubyear{2013}
\maketitle
\label{firstpage}

\begin{abstract}

We study the effects of optically thin radiative cooling on the structure of radiatively inefficient accretion flows (RIAFs).
The flow structure is geometrically thick, and independent of the gas density and cooling, if the cooling time is longer 
than the viscous timescale (i.e., $t_{\rm cool} \gtrsim t_{\rm visc}$). For higher densities, the gas can cool before it can 
accrete and forms the standard 
geometrically thin, optically thick Shakura-Sunyaev disk. For usual cooling processes (such as bremsstrahlung), we 
expect an inner hot flow and an outer thin disk. For a short cooling time the accretion flow separates into two phases:
a radiatively inefficient hot coronal phase and a cold thin disk. We argue that there is an upper limit on the density of the 
hot corona corresponding to a critical value of $t_{\rm cool}/t_{\rm ff} (\sim 10-100)$, the ratio of the cooling time and the free-fall time.
Based on our simulations, we have developed a model for transients observed in black hole X-ray binaries (XRBs). An
XRB in a quiescent hot RIAF state can transition to a cold black-body dominated state because of an increase in the mass 
accretion rate. The transition from a thin disk to a RIAF happens because of mass exhaustion due to accretion; the 
transition happens when the cooling time becomes longer than the viscous time at inner radii.
Since the viscous timescale for a geometrically thin disk is quite long, the high-soft state is expected to be long-lived. The different timescales 
in black hole transients correspond to different physical processes such as viscous evolution, cooling, and free-fall.
Our model captures the overall features of observed state transitions in XRBs.

\end{abstract}


\section{Introduction}

Black holes (BHs) are the simplest astrophysical objects, completely specified by their mass and spin. Black holes are also the 
most powerful sources in the universe because of their strong gravity. Roughly $\sim$10\% of the rest mass energy of the matter 
falling into BHs can be extracted in the form of radiation and jets/outflows. While BHs themselves are quite simple, accretion flows
around them are not. The accretion flows are affected by complex, nonlinear processes such as turbulent angular momentum 
transport, cooling, and radiation (see \citealt{bal98,nar05} for a review).

There are two well-established classes of astrophysical black holes: stellar-mass black holes ($\sim10 \msun$), the end products 
in the evolution of some of the massive stars (see \citealt{woo02} for a review); and supermassive black holes ($\sim 10^6-10^{10} 
\msun$) at the centers of galaxies. Both these classes of BHs show two dominant states of the accretion flow around them: an optically 
thin, geometrically thick, hot radiatively inefficient state (e.g., see \citealt{ich77,ree82}); and an optically thick, geometrically thin black-body disk state 
(\citealt{shakura73}). The quasars observed at high redshifts, with a dominant black-body component, are examples of thin disk accretion 
in supermassive BHs. The quiescent  active galactic nuclei (AGNs) at low redshifts lack thermal disk emission and are modeled as radiatively 
inefficient accretion flows (RIAFs).
Because of shorter timescales in X-ray binaries (XRBs) the same system can be observed to transition from a quiescent state to a black-body state 
and vice-versa (see \citealt{remi06} for a review).

Most accretion flow simulations model the RIAF state because one does not need to worry about cooling and radiation transport 
(e.g., \citealt{stone99,igu03,bec08,pen10}). Another problem in modeling thin disks is the enormous resolution required for a realistic 
$H/R \sim 0.001-0.01$ (ratio of disk height to radius). There are a handful of simulation papers which include some effects of cooling 
on RIAFs (e.g., \citealt{li12,bar12,fra09}). Progress has been made recently to include most of the relevant physics (magnetic fields, radiation, 
cooling, etc.) in numerical simulations (e.g., \citealt{ohs09,ohs11}) but we are still far away from modeling all the processes with 
realistic parameters. 

In this paper we perform detailed hydrodynamic simulations of accretion with varying viscosity prescriptions. The main aim is to study the 
influence of optically thin radiative cooling on the global structure of the accretion flow. While it is widely believed that magnetohydrodynamic (MHD) turbulence associated with the magnetorotational instability is responsible for angular momentum transport and dissipation in accretion flows (\citealt{bal91}), a phenomenological treatment of viscosity allows us to gloss over the complexities (both numerical and physical) of MHD accretion and to focus on the interplay of viscous accretion and cooling. We study the influence of various parameters on the structure of the hot corona and the cold, thin disk.

We confirm the theoretical prediction that a thin disk forms at radii where the cooling time is shorter than the viscous accretion time (\citealt{ree82}). 
We observe an upper limit on the density of the hot corona corresponding to a lower limit on the ratio of the cooling time and the free-fall time, 
$t_{\rm cool}/t_{\rm ff} \sim 10-100$, irrespective of the form  and magnitude of viscosity. This means that there is an upper limit on the amount of mass that can go into
the hot corona; the rest of the mass has to either cool onto the thin disk or be lost in outflows.

Motivated by our idealized simulations, we identify different physical 
processes responsible for the observed characteristics of black-hole binary transients. We argue that the observed transition of the quiescent 
(very low-hard) state to a black-body dominated (high-soft) state via a short-lived high/hard state occurs because of a large increase in the mass 
accretion rate. This transition from hard to soft state can occur at a fast timescale, comparable to the viscous timescale at the inner circularization radius.
The transition from a thin disk back to the low-hard state happens only after sufficient mass has been accreted, so that the cooling time is longer than 
the viscous time everywhere. This occurs at the longest viscous time in the thin disk, which can be years.

The paper is organized as follows. In section 2 we present the hydrodynamic  model and details of numerical implementation. In section 3 we discuss our results
of runs with and without cooling. In section 4 we present a model for global properties of BH XRB transients based on our simulations. In section 5 we conclude 
with astrophysical implications.

\section{Method}

\subsection{Basic Hydrodynamic Equations}
\label{sec:equations}

In our numerical simulations we solve the basic hydrodynamic equations in spherical polar coordinates ($r$, $\theta$, $\phi$) using the ZEUS-MP code \citep{hay06}.
These are, the equation of continuity,
\be
\frac{d \rho}{dt} + \rho {\boldsymbol \nabla} \cdot {\bf v}  = 0,
\label{eq:cont}
\ee
the momentum balance equation,
\be
\rho \frac{d \bf{v}}{dt} = - {\boldsymbol \nabla} P - \rho {\boldsymbol \nabla} \phi + {\boldsymbol \nabla} \cdot {\boldsymbol \sigma},
\label{eq:momentum}
\ee
and the energy equation,
\be
\rho \frac{d(e/\rho)}{dt} =  -P {\boldsymbol \nabla} \cdot \mathbf{v}   + {\boldsymbol \sigma}^{2}/ \mu - n_{e}n_{i}\Lambda(T).
\label{eq:energy}
\ee
Here, $\rho$ is the mass density, $P$ the pressure, $\bf{v}$ $= (v_r, v_{\theta}, v_{\phi})$ the velocity, $e$ the internal energy density ($=3P/2$ corresponding to an 
adiabatic index $\gamma=5/3$; protons are non-relativistic till close to the BH), $\mu = \nu \rho$ the coefficient of viscosity ($\nu$ is the kinematic viscosity), $\boldsymbol \sigma$ the viscous stress tensor, 
$n_e$ and $n_i$ the electron and ion number densities, and $\Lambda(T)$ the cooling function with temperature $T$. The Lagrangian time 
derivative is denoted by $d/dt = \partial/\partial t + \bf{v} \cdot {\boldsymbol \nabla}$. While most of our runs simulate Eqs. (\ref{eq:cont})-(\ref{eq:energy}), in section \ref{sec:conduction} we briefly discuss runs with thermal conduction.

We use the pseudo-Newtonian potential proposed by Paczy\'{n}ski and Wiita (1980), which describes the gravitational field of the central black hole, 
\be
\phi = -\frac{GM}{r-R_{g}},
\label{eq:potential}
\ee
where $M$ is the the black hole mass, $c$ the speed of light, $G$ the gravitational constant, and $R_g (= 2GM/c^2)$ the Schwarzschild radius.  

We include a bremsstrahlung cooling term in the energy equation (last term on the right hand side of Eq. \ref{eq:energy}). We use the cooling function $\Lambda(T)$ given by equation (12) of \cite{sharma}, which has been adopted from \cite{sutherland} for mean molecular weights corresponding to $\mu=0.62$ and $\mu_e=1.18$. We note that in the present work we ignore cooling due to synchrotron and inverse Compton processes, which although important in certain accretion regimes (e.g., \citealt{raj10,dra13}), are difficult to model. 
Both synchrotron and Compton cooling terms are expected to be proportional to $n^2$ (square of number density) because both the magnetic and radiation energy densities are expected to scale with the mass accretion rate. Thus, including these should give results qualitatively similar to bremsstrahlung cooling. A further simplification that we make is to use a single fluid model;  ions in RIAFs are expected to much hotter than electrons (\citealt{ree82}). We will work to overcome these shortcomings in future.

The second term on the right hand side of Eq. (\ref{eq:energy}) represents heating due to viscous dissipation. 
MHD simulations~(e.g., Table 2 in \citealt{haw95}; \citealt{bra95}) show that the $r\phi$ component of the magnetic stress is the most dominant component of the stress tensor. Although we do not include magnetic fields in this work, in order to mimic its effects, we assume that only the $r \phi$ component of the viscous stress tensor $\bf{\sigma}$ is non-zero, such that:
\begin{equation}
\sigma_{r \phi} = \sigma_{\phi r} = \mu r \frac{\partial}{\partial r} \left(\frac{v_{\phi}}{r}\right).
\label{eq:trphi}
\end{equation}
This anisotropic stress behaves very differently from an isotropic viscous stress, and only damps radial variations in the angular velocity.

We choose the fiducial form of the kinematic viscosity in our runs to be $\nu \propto r^{1/2}$, guided by the existing theoretical predictions. In the standard `$\alpha-$disk' model of a thin disk, $\nu = \alpha c_{s} H$, where $\alpha$ is the Shakura-Sunyaev viscosity parameter, $c_{s}$ the sound speed and $H$ the disk scale height \citep{shakura73}. Now, $H \sim c_s/\Omega$, where $\Omega$ is the angular velocity, and we have, $\nu \approx \alpha H^{2} \Omega \approx \alpha (H/r)^{2} \Omega r^2$. For a Keplerian flow, $\Omega \propto r^{-3/2}$ and hence, $\nu$ scales with $r$ as $\nu \propto r^{1/2}$, provided $H/r$ is constant, which is indeed the case for a Shakura-Sunyaev disk. For comparison we also try $\nu = {\rm constant}$ and $\nu \propto \rho$. The kinematic viscosity for the fiducial case is given by 
\be
\nu = 4.5\times 10^{-4} \alpha_{0.01} \Omega_0 R_0^2 \left(\frac{r}{R_0}\right)^{1/2},
\label{eq:nu}
\ee
where $\alpha_{0.01}=\alpha/0.01$, $R_0$ is the center of the torus where the initial density maximum is located (see section 2.3) and $\Omega_0$ is the angular velocity at $R_0$; most runs use $\alpha=0.01$ but some runs which are evolved for long times use $\alpha=0.1$. Therefore, the viscous time is $\approx 1600 \alpha_{0.01}^{-1}$ times the free-fall time (see Eqs. \ref{eq:tvisc} \& \ref{eq:tdyn}). The magnitude of the constant viscosity (for $\nu= {\rm constant}$ case) is obtained by evaluating $\nu$ in Eq. 6 at $R_0$. Similarly, the normalization for $\nu \propto \rho$ case is chosen such that the viscous timescale at $R_0$ is the same as in $\nu \propto r^{1/2}$ case.

Note that, in absence of cooling, Eqs. 1-3 are independent of the density normalization (since $P \propto \rho$ and ${\boldsymbol \sigma} \propto \mu \propto \rho$, provided the kinematic viscosity $\nu$ is not a function of density). Thus, the accretion flow structure in absence of cooling is identical if we normalize the density by its initial maximum value (this can be seen from Figures 1(e) and 3(a) which show density at late times for RIAFs with different initial densities). Furthermore, the only spatial scale in Eqs. 1-4 is the Schwarzschild radius $R_g$, which is proportional to the black hole mass (e.g., \citealt{raj10}). Thus, if we scale all the spatial scales with $R_g$, velocities with $c$, time with $R_g/c$, density with the initial maximum value $\rho_0$, $P$ and $e$ by $\rho_0 c^2$, $\nu$ by $c R_g$, then in absence of cooling (i.e., for RIAFs) there is no explicit dependence on the black hole mass in Eqs. 1-4. Therefore, our results can be scaled to a wide range of systems --- from accretion flows around stellar mass black holes, namely X-ray binaries (XRBs), to those around supermassive black holes, namely active galactic nuclei (AGNs).

\subsection{Key Timescales and RIAFs}
 
Here we list some of the important timescales involved in accretion flows with cooling.
The viscous timescale of the accretion flow, which is the time taken by the matter to accrete onto the black hole, is given by
\be
t_{\rm visc} = \frac{r^2}{\nu}.
\label{eq:tvisc}
\ee
We define the free fall timescale of the flow to be
\begin{equation}
t_{\rm ff} \equiv \sqrt[]{\frac{2r^{3}}{GM}}.
\label{eq:tdyn}
\end{equation}
The cooling timescale of the flow is given by
\begin{equation}
t_{\rm cool} = \frac{e}{\dot{e}}, 
\label{eq:tcool}
\end{equation}
where $\dot{e}=n_{e}n_{i}\Lambda(T)$ is the cooling rate. If $k_B$ is Boltzmann's constant and $n$ denotes the total number density then $e =3nk_BT/2$ and $\dot{e} \approx -n^2 \Lambda(T)$, and hence $t_{\rm cool} \propto n^{-1}$. Thus the cooling time is shorter for a denser optically thin accretion flow.

If the cooling time is shorter than the viscous time (i.e., $t_{\rm cool} \lesssim t_{\rm visc}$) in RIAFs the accretion flow separates into two phases, a coronal phase with long cooling times and a dense optically-thick, geometrically thin disk. Both these phases are in thermal equilibrium: viscous heating balances black-body radiation in the thin disk; viscous heating balances optically thin cooling and adiabatic expansion in the coronal phase. A dense accretion flow develops these two phases because an optically thin flow which cools efficiently is globally unstable (e.g., \citealt{pir78}). The two-phase corona-disk is globally stable. If cooling in the coronal phase dominates heating, excess mass can condense onto the thin disk and equilibrium is achieved. Similarly if heating in the corona dominates, excess energy goes into evaporating matter from the thin disk and in launching outflows. Although the corona is in global thermal equilibrium, it can become locally thermally unstable for short  cooling times, and excess mass from the  corona can condense into cold blobs which fall on the thin disk (see section \ref{sec:corona} and \citealt{icm}).

In our simulations we do not include radiative transport and hence are unable to simulate a realistic standard thin disk. Note that, as the RIAF cools, it becomes dense and radiation does not simply escape, but diffuses with a short mean free path. The gas becomes optically thick before cooling to $10^4$ K (which is the temperature of the stable phase for our cooling function; cold thin disk exists at this temperature in our simulations). Therefore, in reality, the thin disks should be hotter and of lower density compared to those in our simulations. We note that our thin disks are severely unresolved because photon transport (e.g., \citealt{ohs11}) or magnetic fields (e.g., \citealt{mac06}) which can provide vertical support for the cooling gas, are absent.  The present paper focuses on the structure of RIAFs and the conditions for the formation of a thin disk, not on its detailed structure. Realistic thin disk simulations with radiation transport will be carried out in future.

\begin{figure*}
\centering
\includegraphics[scale=0.5]{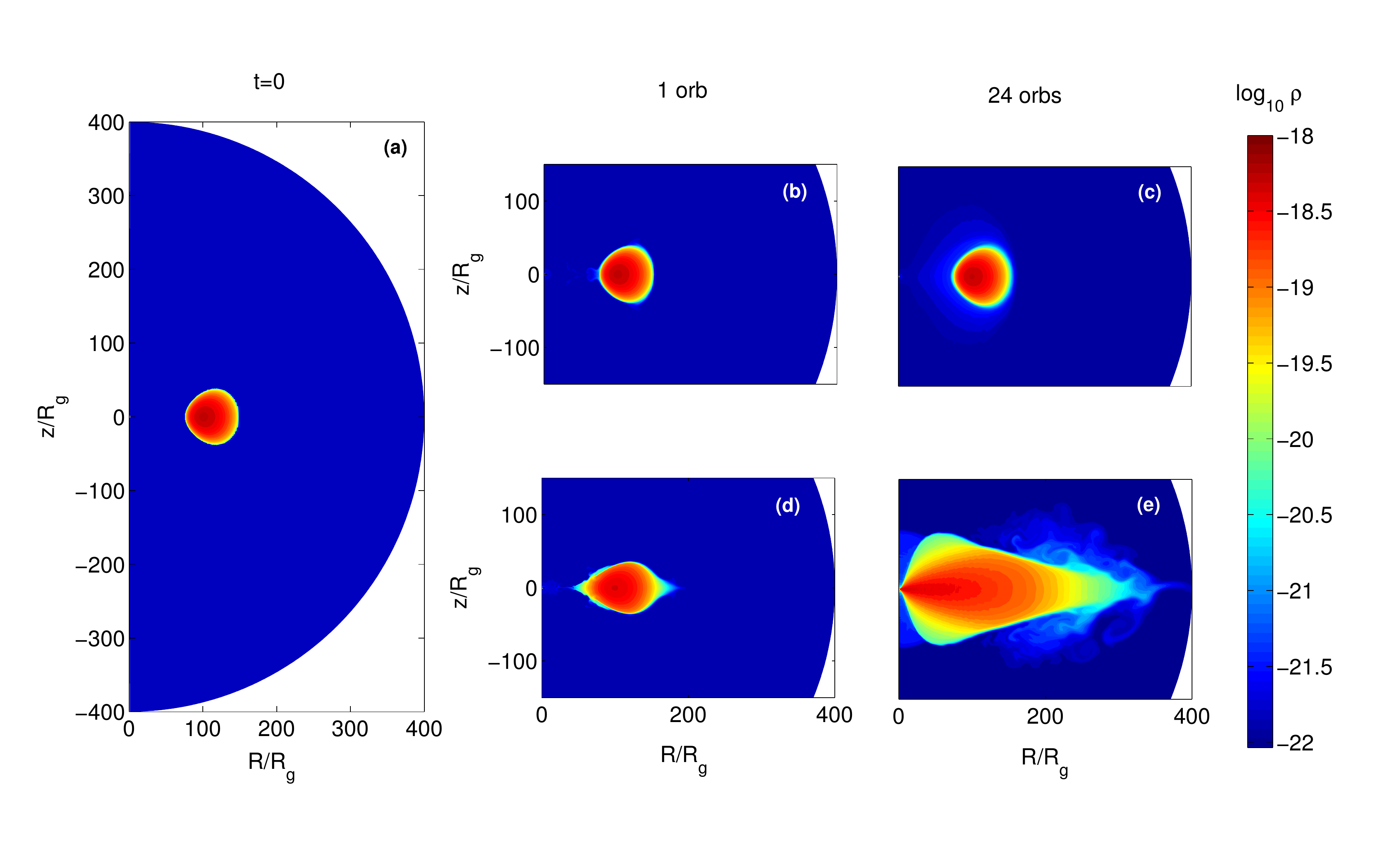}
\caption{(a) Initial density contour plot for the run with $\rho_0=10^6m_p$ $\rm g~cm^{-3}$. Panels (b,c) show evolution without viscosity and panels (d,e) with viscosity $\nu \propto r^{1/2}$ and $\alpha=0.01$ (see Eq. \ref{eq:nu}). The grid resolution is $512^2$.}
\label{fig:novisc}
\end{figure*}

\subsection{Initial Conditions}
\label{sec:initialcond}

Our initial condition consists of a dense torus having a constant specific angular momentum in dynamical equilibrium. The torus is embedded in a medium having very low density and is non-rotating. The pressure and density in the torus satisfy the polytropic equation of state $P = K \rho^{\gamma}$, $K$ being a constant. Then following \citealt{pp1984} and \citealt{stone99} we determine the the initial structure of the torus, for our potential $\phi$, as:
\be
\frac{P}{\rho} = \frac{GM}{(n+1)R_0}\left[\frac{R_0}{r-R_g} - \frac{1}{2}\frac{R_{0}^{4}}{[(R_0-R_g)R]^{2}} - \frac{1}{2d}\right],
\label{eq:papaloz}
\ee
where, $R=r\sin \theta$, is the cylindrical radial coordinate; $R_0$ is the equatorial distance of the center of the torus (where it has maximum density) from the black hole; $d$ ($>1$) is the distortion parameter which determines the shape  and size of the torus $-$ if $d \approx 1$, then the torus has a small, nearly circular cross-section and as $d$ becomes much greater than $1$, the cross-section deviates from a circular one and the torus becomes fatter; $n=1/(\gamma -1)$ is the polytropic index. For most of our runs the distortion parameter is taken to be $d=1.125$. The density of the torus is maximum ($\rho = \rho_0$) at $r = R_0$ and $\theta = 90^\circ$, using which one can write down the expression for the constant $K$:
\be
K = \frac{GM}{(n+1)R_0 \rho_0^{\gamma -1}}\left[\frac{R_0}{R_0-R_g} - \frac{1}{2}\frac{R_{0}^{2}}{(R_0-R_g)^{2}} - \frac{1}{2d}\right].
\label{eq:K}
\ee
Thus, knowing $\rho_0$ and $d$ one can easily determine $K$. 

The initial density of the embedding ambient medium is chosen to be $\rho_{\rm amb} = 10^{-4} \rho_0$, which is too small to affect our results. The mass of the black hole is taken to be that of Sgr $A^{*}$, i.e., $M = 4\times 10^6 M_{\odot}$ \citep{ghez08}. We scale most of our results for both XRBs and AGNs since the governing equations can be scaled for different BH masses.

\subsection{Numerical Setup}

We carry out two-dimensional, axisymmetric, hydrodynamic simulations in spherical $(r,\theta,\phi)$ geometry, with the computational grid extending from an inner boundary $R_{\rm min}=1.75 R_g$ to an outer boundary $R_{\rm max} = 400 R_g$. A logarithmic grid is used along the radial coordinate, while a uniform grid is used along the $\theta$ direction, where $0 \leq \theta \leq \pi$. Although there is no variation along the $\phi$ direction, it ranges from $0 \leq \phi \leq 2\pi$.

As for the boundary conditions, we use inflow condition at the inner radial boundary, such that mass is allowed to flow into the black hole but not out of it. We apply outflow condition at the outer radial boundary, such that mass is allowed to flow out of the grid but not into it. In addition, the stress $\sigma_{r \phi}$ is set to zero at the inner radial boundary. 

Our standard resolution is $N_r=N_\theta=512$, where $N_r$ and $N_\theta$ denote the number of grid points in the $r$ and $\theta$ directions, respectively. For some simulations, which are evolved for many viscous times at large radii and hence require a relatively longer computational time, we have used low-resolution models with $N_r=N_\theta=256$ or $N_r=N_\theta=128$.

In some of the high resolution runs with cooling the density can become unusually low at some isolated grid points. In order to avoid this and the associated numerical problems, we have used appropriate density and internal energy floors in our simulations, which however, do not affect the overall accretion flow. 

\section{Results}

\begin{table}
\caption{Important runs with cooling}
\begin{tabular}{cccc}
\hline
\hline
{Runs} & $\alpha$ & $\rho_0/m_p$ ($\rm cm^{-3}$) & Resolution \\
\hline
A.01n9 & $0.01$ & $10^9$ &  $512^2$ \\
A.01n10 & $0.01$ & $10^{10}$ & $512^2$ \\
A.01n11 & $0.01$ & $10^{11}$ & $512^2$ \\
A.1n9 & $0.1$ & $10^{9}$ & $256^2$ \\
A.1n10 & $0.1$ & $10^{10}$ & $512^2$ \\
A.1n11 & $0.1$ & $10^{11}$ & $512^2$\\
\hline
\end{tabular}\\
\end{table}

In addition to runs with cooling which are the focus of this paper, we carry out a few runs without cooling to study the pure RIAF state and its dependence on various parameters. Since accretion flows without cooling have already been extensively analyzed in past (e.g., \citealt{igu03,stone99,pro03}), we do not discuss them in detail in this work. 
The important runs with cooling are summarized in Table 1. 

\subsection{Runs without Cooling}

Here we discuss briefly the pure RIAF runs where cooling is switched off (i.e., the last term in Eq. (\ref{eq:energy}) is dropped). As discussed earlier, in this regime the density just scales with the initial maximum density, and the flow is self similar, if the spatial and temporal scales are normalized by the BH mass.

For pure RIAF runs the maximum density of the torus is chosen to be $\rho_0 = 10^6 m_p$ $\rm g~cm^{-3}$ ($m_p$ is the mass of the proton) at the torus center $R_0 = 100 R_g$. The density is loosely motivated by Sgr $A^*$ observations (\citealt{bag03}). The torus is evolved for about 24 orbits at $r=R_0$ to ensure that a steady state is reached (at least at inner radii). Figure~\ref{fig:novisc}(a) shows the density contour plot for the initial condition ($t=0$). Since viscosity is responsible for accretion, in absence of viscosity (i.e., for $\nu=0$) the torus should not evolve and must retain its initial condition. Figures \ref{fig:novisc}(b,c) which show the density at late times indicate that this is indeed the case; a very low density film forms around the initial torus at late times as the equilibrium adjusts slightly. This case shows that numerical viscosity is negligible and accretion happens only because of  explicit viscosity in Eq. \ref{eq:momentum}.

The run with $\nu \propto r^{1/2}$ shows viscous accretion and the formation of a geometrically thick RIAF at small radii at late times (see Figs. \ref{fig:novisc}(d,e)). The matter tapers into a tail like structure in the outer region, which is necessary because some matter has to move out to carry the angular momentum outwards. The inner bulge is a low density ($\rho \gtrsim 10^{-20}$ $\rm g~cm^{-3}$) corona. We use the term `corona' quite loosely for the hot phase of the accretion flow with a long cooling time. Note that this `corona' is formed as a result of accretion and should not be confused with the very low density ambient medium ($\rho_{\rm amb} = 10^{-4} \rho_0$), which is required for numerical reasons.

In order to investigate how sensitive the accretion flow is to different parameters, we have carried out several simulations by taking different initial values of the distortion parameter $d$, the location of the torus center $R_0$, and different functional forms of $\nu$. 
First, the initial torus was placed at different locations from the black hole, namely at $R_0 = 50, 100~ {\rm and}~ 200 R_g$ (note that $d=1.125$ and $\nu \propto r^{1/2}$ for all three cases). For all the cases, the density contours in the inner region of the accretion flow in steady state, were found to be very similar in overall shape. Next, different initial cross-sectional areas of the torus were chosen, namely $d=1.05$, which correspond to a smaller nearly circular torus, $d=1.3$, which consists of a fatter torus deviating from a circular cross section, and $d=1.125$, which correspond to a torus midway between the two extreme cases (note that $R_0=100 R_g$ and $\nu \propto r^{1/2}$ for all three cases). Again all the cases were found to exhibit very similar density contours in steady state. These results indicate that the {\em final steady state is not too sensitive to the initial conditions} and we conclude that the accretion flow is independent of the initial distribution of matter around the black hole, and the way the black hole is fed, as long as the circularization radius ($R_0$) is far away from the Schwarzschild radius.

We have also used different functional forms of the kinematic viscosity, namely, $\nu = \rm constant$ and $\nu \propto \rho$ (note that $d=1.125$ and $R_0=100 R_g$). In order to assess the dependence of the accretion flow on $\nu$, one can arrive at a rough scaling of $\nu$ and density with radius from the conservation of angular momentum (c. f. Eq. 2.8 of \citealt{pringle81}). In steady state this implies that the viscous and the advective fluxes in the equatorial plane satisfy
\begin{equation}
\label{eq:nu_rho}
\nu \rho H R^3 \frac{d \Omega}{dR} \approx  \frac{\dot{M}\Omega R^2}{2\pi},
\end{equation}
where the accretion rate $\dot{M}=\rm constant$. Now, on assuming that the rotational velocity is nearly Keplerian, the above equation leads to $\nu \rho H = \rm constant$. Thus {\em the structure of a RIAF depends on the form of the kinematic viscosity}. The results from these two runs ($\nu=$constant and $\nu \propto \rho$) look very similar, as explained in \cite{stone99}. They differ from the $\nu\propto r^{1/2}$ case in that, they show a more pronounced corona at inner radii. Our results with varying viscosity ($\nu$) are consistent with Eq. \ref{eq:nu_rho}.

\begin{figure*}
\centering
\includegraphics[scale=0.5]{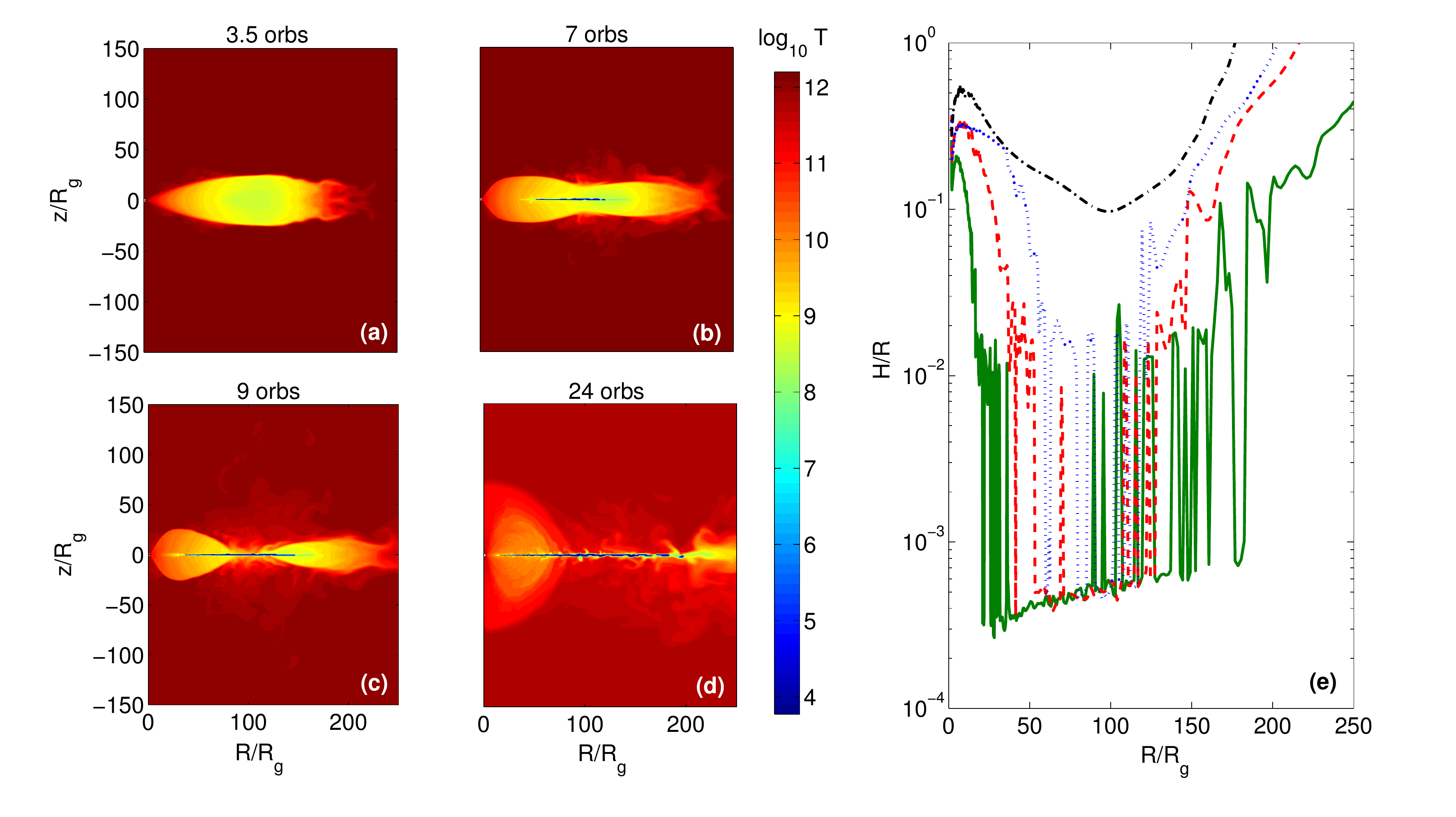}
\caption{(a) (b) (c) (d) The time evolution of temperature contour plots for the fiducial cooling run with $\rho_0=10^{10} m_p$ $\rm g~cm^{-3}$ and $\alpha=0.01$. (e) The variation of the mid-plane aspect ratio ($H/R$) with radius ($R/R_g$). The black (dot-dashed), blue (dotted), red (dashed), and green (solid) lines correspond to the snapshots (a), (b), (c), and (d) respectively. The grid resolution is $512^2$.}
\label{fig:brems1e10}
\end{figure*}

\begin{figure*}
\centering
\includegraphics[scale=0.5]{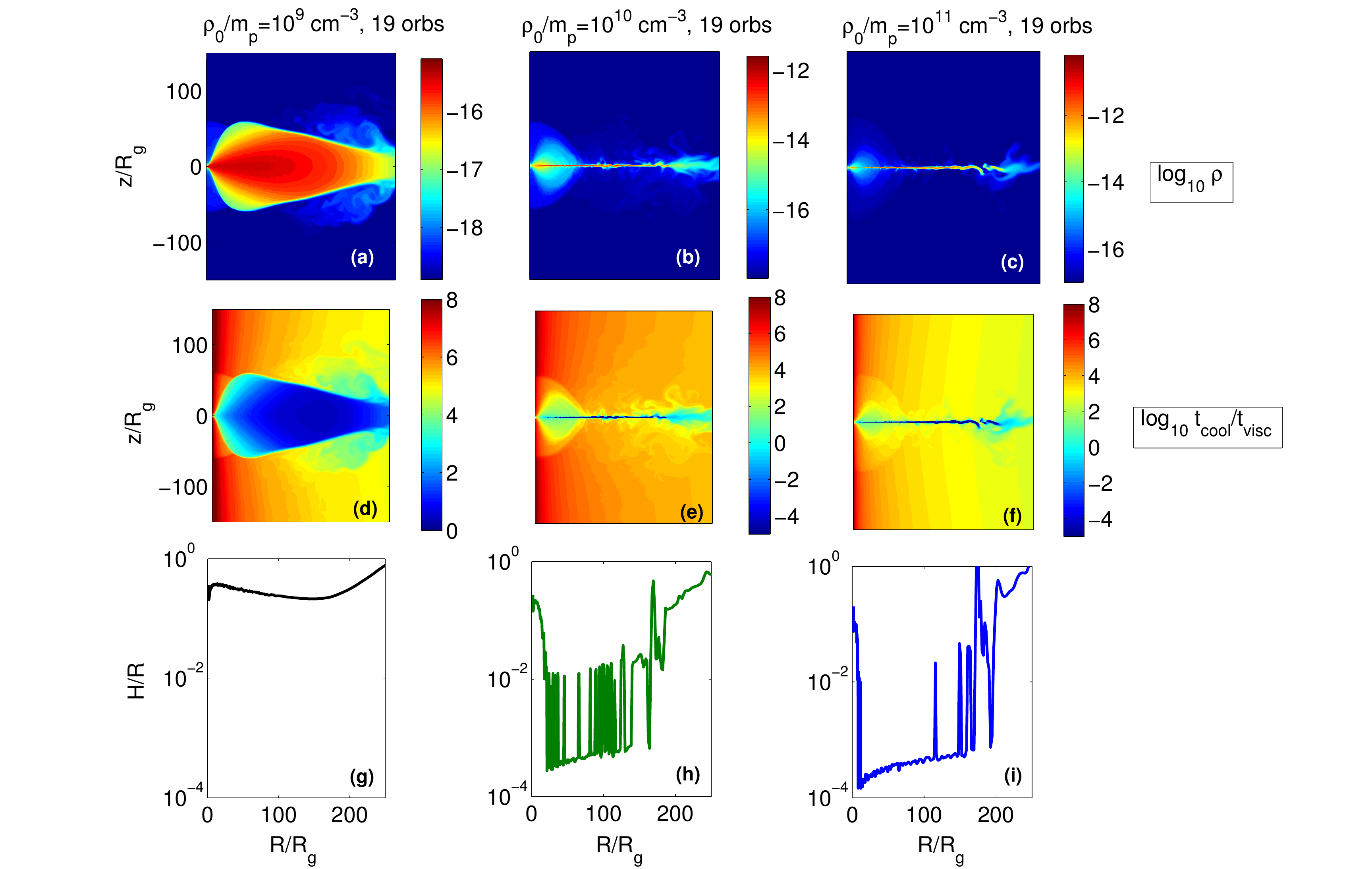}
\caption{(a) (b) (c) Top row - shows the density contour plots in the steady state for runs with three different initial densities. (d) (e) (f) Middle row - shows the contour plots of the ratio of cooling to viscous time ($t_{\rm cool}/t_{\rm visc}$). (g) (h) (i) Bottom row - shows the variation of $H/R$ with $R/R_g$. The grid resolution is $512^2$.}
\label{fig:cool3x3}
\end{figure*}

\begin{figure*}
\includegraphics[scale=0.5]{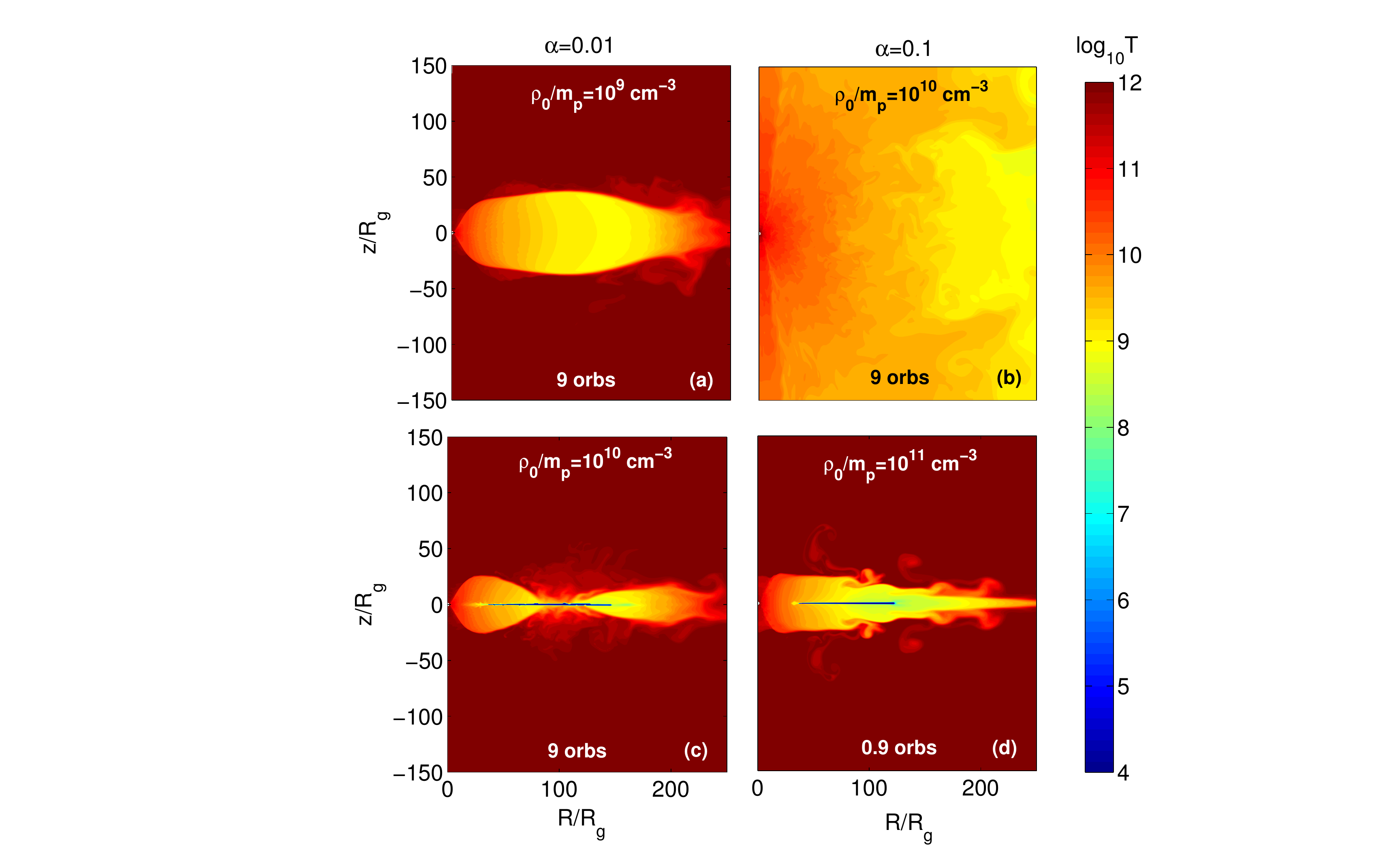}
\caption{Temperature contour plots  for different runs illustrating the importance of the ratio $t_{\rm cool}/t_{\rm visc}$: (a) $\alpha=0.01$, $\rho_0=10^{9} m_p$ $\rm g~cm^{-3}$ (b) $\alpha=0.1$, $\rho_0=10^{10} m_p$ $\rm g~cm^{-3}$ (c) $\alpha=0.01$, $\rho_0=10^{10} m_p$ $\rm g~cm^{-3}$ (d) $\alpha=0.1$, $\rho_0=10^{11} m_p$ $\rm g~cm^{-3}$. The grid resolution is $512^2$.}
\label{fig:brems1e10_diffalpha}
\end{figure*}

\subsection{Runs with Cooling}
\label{sec:cooling}

In this subsection we discuss in detail the effects of bremsstrahlung cooling on the accretion flow. Simulations without cooling are invariant to any change in $\rho_0$, but the runs with cooling are not. Low density runs are essentially RIAFs throughout (even with cooling), but an increase in density eventually leads to the condensation of the hot gas and the formation of a cold thin disk. Ideally the accretion rate $\dot{M}$ (or even better $\dot{M}/\dot{M}_{\rm Edd}$, where $\dot{M}_{\rm Edd}$ is the Eddington rate) should be chosen as the parameter to study this transition, but $\rho_0$ is a useful proxy.

\subsubsection{The Fiducial Cooling Run}

Figures 2(a)-(d) show the variation of the two-dimensional temperature contour plots with time for the accretion flow having an initial maximum density $\rho_0=10^{10} m_p$ $\rm g~cm^{-3}$ and viscosity parameter $\alpha=0.01$ (see Eq. \ref{eq:nu}; Run A.01n10 in Table 1). The accretion flow starts off as a geometrically thick, hot flow. It gradually develops a peanut-like shape after about 7 orbits (roughly the cooling time at $R_0$) and also shows the onset of cold high density gas in the mid-plane. After 9 orbits the low density corona displays a pinch at about $100R_g$, and a more extended cold, thin disk is formed. Once steady state is attained, one sees the clear formation of an inner low density, hot ($T \sim 10^{10-11}$ K), geometrically thick corona and a geometrically thin, cold ($T \sim 10^{4-5}$ K) disk which extends till very close to the black hole. 

Since the cooling time is inversely proportional to the density, a higher density (or equivalently a higher mass accretion rate) corresponds to a shorter cooling time and an increase in the radiative efficiency of the flow. Thus, we conclude that the initial density of the torus in this case is high enough to trigger the onset of bremsstrahlung cooling, which eventually condenses the hot flow to form a geometrically thin cold disk. Some corona models require conductive heating of the upper layers of the thin disk to form the corona via evaporation  (e.g., \citealt{mey94}; they also include viscous heating but it seems to be subdominant in their models). Our corona is maintained by viscous heating; in reality, magnetic dissipation and evaporation due to thermal conduction (magnetized thermal conduction is quite non-trivial to model faithfully) should  also play a significant role ( see section \ref{sec:conduction}). Irrespective of the source of heating, the corona exists in a state of rough dynamical and thermal equilibrium, where maximum density is governed by the physics of local thermal instability in presence of gravity (see also section \ref{sec:corona}).

The transition to a thin disk is further illustrated in Figure 2(e), which shows the variation of the disk height to radius ratio ($H/R$; a measure of geometrical thickness of the flow) as a function of radius ($R/R_g$). The disk scale height $H \approx c_s/\Omega$ is calculated using the mid-plane ($\theta=\pi/2$) value of  $c_s \approx \sqrt[]{P/\rho}$ and $\Omega \approx \sqrt[]{GM/R^3}$. The black (dot-dashed), blue (dotted), red (dashed) and green (solid) lines show the $H/R$ of the accretion flow at 3.5, 7, 9 and 24 orbits, respectively. Initially the flow is geometrically thick throughout, with a high $H/R \gtrsim 0.1$. As accretion continues,  $H/R$ starts dropping near the density maximum (around $100R_g$), leading to the formation of a geometrically thin disk. We note that the innermost region is always geometrically thick with $H/R \gtrsim 0.1$, however, as time progresses, the geometrically thin disk with $H/R \sim 5 \times 10^{-4}$ moves closer in to the black hole (up to about $20 R_g$). Note that Figure 2(e) shows spikes in $H/R$ in the region of the cold thin disk because there is hot gas at some of the grids in the equatorial plane (at $z=0$). This increases the estimate of $H/R$ locally. Also note that the extreme outer region is not a thin disk because the low-density hot gas has a long cooling time.

The accretion flow is expected to have an inner RIAF and an outer cold disk because the cooling time, $t_{\rm cool} \propto T^{1/2}/n$, is expected to decrease outward faster than the viscous time, $t_{\rm visc} = r^2/\nu$. As we discuss in the next sections, this ratio determines whether a thin disk forms at a given radius.

\subsubsection{Dependence on Initial Density}

The fact that a minimum threshold density is required to trigger efficient bremsstrahlung cooling is better illustrated in Figure 3, which compares results from simulations having three different initial densities. Figure 3 shows the steady state contour plots of density (top row), the ratio $t_{\rm cool}/t_{\rm visc}$ (middle row), and the variation of $H/R$ with radius (bottom row), for accretion flows with different initial density normalizations: $\rho_0=10^{9} m_p$ $\rm g~cm^{-3}$ (left column; Run A.01n9), $\rho_0=10^{10} m_p$ $\rm g~cm^{-3}$ (middle column; Run A.01n10) and $\rho_0=10^{11} m_p$ $\rm g~cm^{-3}$ (right column; Run A.01n11). Note that all the three cases have the same viscosity parameter $\alpha$ (= 0.01; see Eq. \ref{eq:nu}), i.e., the same viscous time, but  different cooling times due to the difference in the density of the accreting matter. 

We observe that the accretion flow with the lowest initial density $\rho_0=10^{9} m_p$ $\rm g~cm^{-3}$ in Figure 3(a) is throughout geometrically thick with $H/R \gtrsim 0.1$ (Figure 3(g)) and resembles the flow without cooling in Figure 1(e). The entire flow can be classified as a hot, low density corona which is radiatively inefficient. However, as $\rho_0$ increases to $10^{10} m_p$ $\rm g~cm^{-3}$ (Figure 3(b)) the hot corona is restricted to the innermost radii and there is appearance of a cold, geometrically thin disk with $H/R \sim 5\times 10^{-4}$ (Figure 3(h)). The time evolution of this run is already discussed previously (see Figure \ref{fig:brems1e10}). For $\rho_0 = 10^{11} m_p$ $\rm g~cm^{-3}$ the cold disk extends very close to the black hole (up to $\lesssim 10R_g$) (Figure 3(i)) and the inner coronal bulge has diminished significantly relative to the disk (Figure 3(c)). Thus, $\rho_0=10^{10} m_p$ $\rm g~cm^{-3}$ seems like a threshold density, which marks the transition from a geometrically thick hot flow to a geometrically thin cold disk. 

The condition for the formation of a geometrically thin disk is governed by the ratio of the cooling time and the viscous time ($t_{\rm cool}/t_{\rm visc}$). Figure 3(d) shows that $t_{\rm cool}/t_{\rm visc}$ is everywhere greater than 1. Thus the flow is dominated by viscous effects and cooling does not affect the flow because the gas accretes before it can cool. As the initial density ($\rho_0$) increases the cooling time becomes shorter and cooling starts to become important. Figure 3(d) shows that the ratio $t_{\rm cool}/t_{\rm visc}$ will become smaller than unity close to the dense torus if the initial density is increased by a factor of ten. Indeed, the run with $\rho_0=10^{10} m_p$ g cm$^{-3}$ shows a thin disk with $t_{\rm cool}/t_{\rm visc} \ll 1$ beyond $20 R_g$ in the steady state and the inner RIAF has $t_{\rm cool}/t_{\rm visc} \gtrsim 1$ (Figure 3(e)). A similar conclusion is drawn from Figures 3(c) and (f) for the case with $\rho_0 = 10^{11} m_p$ $\rm g~cm^{-3}$, which is a completely cooling dominated flow, resulting in a thin accretion disk extending till within $10R_g$. The $H/R$ ratio in the thin disk for the two high density cases is similar ($\sim 5 \times 10^{-4}$). As mentioned earlier, this ratio for the cold thin disk is not physically consistent because vertical radiative transport is ignored in our simulations. 

\subsubsection{Dependence on $\alpha$ Parameter}

\begin{figure*}
\centering
\includegraphics[scale=0.5]{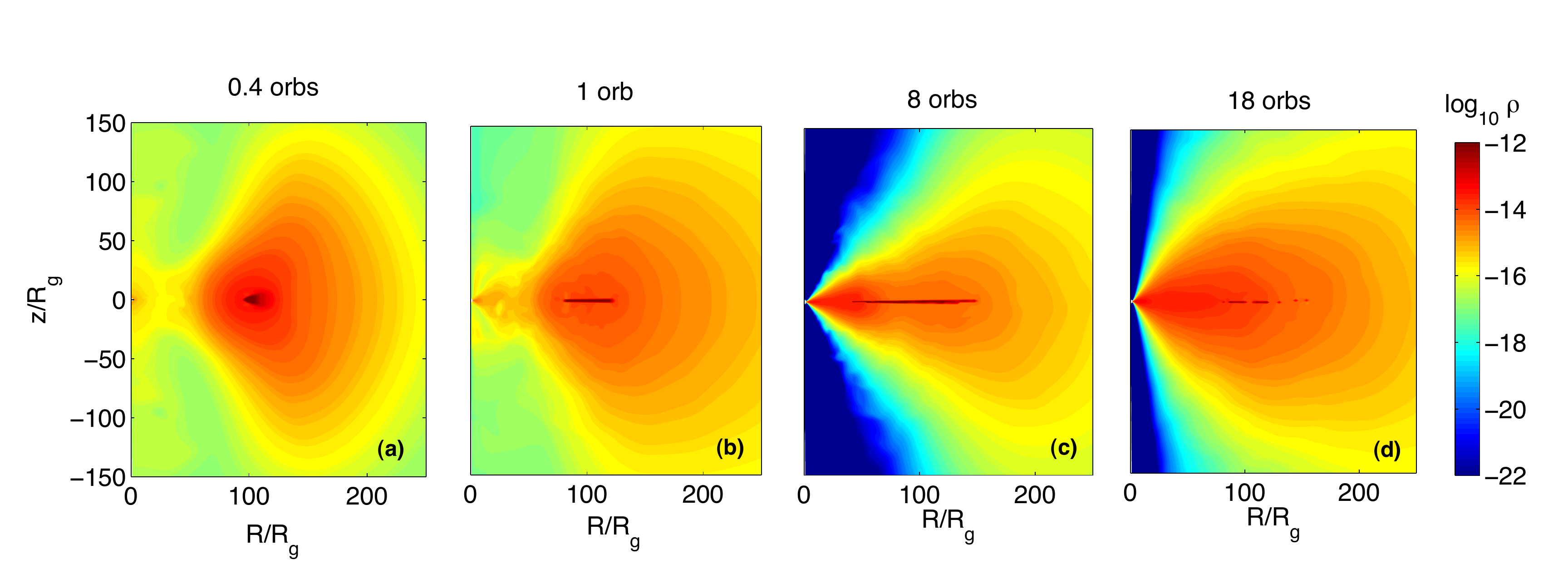}
\caption{(a) (b) (c) (d) Density contour plots at different times for the cooling run with $\rho_0=10^{11} m_p$ $\rm g~cm^{-3}$ and with conduction. Note that at early times (a,b) large  conductive heating pushes matter close to the poles; eventually this excess mass at the poles is accreted and we get the usual low density funnel at the poles (c,d). The grid resolution is $256^2$.}
\label{fig:conduction}
\end{figure*}

So far we have only considered flows with different initial densities but with the same viscosity parameter $\alpha$ (see Eq. \ref{eq:nu}). If the criterion for the formation of a cold thin disk is $t_{\rm cool}/t_{\rm visc} \lesssim 1$ then increasing $\alpha$ (i.e., decreasing $t_{\rm visc}$) should result in a higher initial density for which the transition of a RIAF to a thin disk happens. To test this, we study two runs with $\alpha=0.1$ (ten times the fiducial value), and initial densities $\rho_0=10^{10} m_p$ g cm$^{-3}$ and  $\rho_0=10^{11} m_p$ g cm$^{-3}$. Since the viscous time is ten times shorter than the fiducial case, $t_{\rm cool}/t_{\rm visc}$ in steady state for $\rho_0=10^{10} m_p$ g cm$^{-3}$ and $\alpha=0.1$ run is expected to be similar to the run with $\rho_0=10^{9} m_p$ g cm$^{-3}$ and $\alpha=0.01$ (Fig. 3(d)). Therefore, we do not expect the $\rho_0=10^{10} m_p$ g cm$^{-3}$ run with $\alpha=0.1$ to show a thin disk in the steady state. This is indeed the case as can be seen from Figure 4(b).

Figure 4 shows the two-dimensional temperature contour plots of runs with different viscosity parameters ($\alpha$s) and initial density normalizations ($\rho_0$s). The left column corresponds to the runs with $\alpha=0.01$, while the right column corresponds to those with $\alpha=0.1$. Interestingly, although Figures 4(b) and (c) correspond to the same $\rho_0$, they look absolutely different. After 9 orbits the run with $\rho_0=10^{10} m_p$ g cm$^{-3}$ and $\alpha=0.01$ (our fiducial cooling run) shows the condensation of a cold geometrically thin disk with $T\sim 10^4$ K, while the run with the same density but with $\alpha=0.1$ shows a pure RIAF. Figure 4(d) shows the formation of a cold thin disk for a ten times higher density run ($\rho_0=10^{11} m_p$ g cm$^{-3}$) with $\alpha=0.1$. Although Figures 4(c) and (d) are plotted at 9 and 0.9 orbits respectively, they correspond to the same viscous time and hence the thin disks look quite similar in the two cases. Similarly, while the RIAF temperature contour plots in Figure 4(a) and 4(b) are plotted at the same time, their viscous times differ by a factor of 10. Therefore the higher viscosity run has a more extensive corona; we expect a similar flow structure for the $\alpha=0.01$, $\rho_0=10^9 m_p$ g cm$^{-3}$ run at 90 orbits.
 
\subsection{Effects of Thermal Conduction}
\label{sec:conduction}

Thermal conduction is expected to modify the accretion flow, mainly by evaporating the thin disk and by making the corona much more widespread (e.g., \citealt{mey94}). In this section we briefly consider the effects of thermal conduction on some of our simulations with cooling. We defer a detailed study of thermal conduction to future. 

For our runs with conduction, a thermal conduction term $-{\bf \nabla \cdot Q}$ is added to the right hand side of the internal energy equation (Eq. \ref{eq:energy}), where the heat flux is given by ${\bf Q } = -\kappa {\bf \nabla}T$, with diffusivity given by the Spitzer form $\chi = \kappa T/P = 4.7 \times 10^9 T^{5/2}/n$ cm$^2$ s$^{-1}$, where $n=n_e+n_i$ is the total number density.  We limit the conductivity such that the conduction timescale ($r^2/\chi$) is everywhere longer than 0.1 times the local dynamical time ($[r^3/GM]^{1/2}$). This reduces the conductivity in the ambient medium and prevents the conduction timestep from being excessively short. It also roughly mimics the effect of saturated conduction in extremely hot plasmas (e.g., see \citealt{sha08}). Since runs with conduction are quite expensive, we use the resolution of $256^2$.

We find that our fiducial cooling run with $\rho_0=10^{10} m_p$ g cm$^{-3}$ does not show the formation of a cold, thin disk with conduction. In this case, thermal conduction is able to diffuse the dense torus by conductively heating it, thereby reducing the density of the accretion flow (thermal conduction generically reduces the mass accretion rate; e.g., \citealt{sha08}). Therefore, the ratio $t_{\rm cool}/t_{\rm visc} \gtrsim 1$ everywhere and no thin disk can condense out of the accretion flow. 

Figure 5 shows the density snapshots of our $\rho_0=10^{11} m_p$ g cm$^{-3}$ run with conduction at different times. In this case, the dense core of the initial torus is able to cool to low temperatures (Fig. 5(b)) before being diffused by thermal conduction. The corona is much more extensive because of conductive spreading (compare with Fig. 3(c)). The small scale features, seen in runs without conduction, are wiped out because of thermal conduction. The cold disk condenses at the density maximum (Fig. 5(a)), moves inward viscously (Fig. 5(b)), and forms a quasi-steady disk (Fig. 5(c)). Eventually, the inner thin disk moves outward and is evaporated (Fig. 5(d)), and only a hot accretion flow remains. The thin disk no longer exists because thermal conduction has spread out (and diluted) the corona over such a large scale that $t_{\rm cool}/t_{\rm visc} > 1$ everywhere (outflows can also do this by removing mass from the accretion flow). In runs without conduction the cold disc dissipates at a much longer viscous timescale (see section 3.4.2). 
 
To summarize, thermal conduction has a large impact on the accretion flow, in particular the corona is extended and the mass accretion rate is reduced because of conduction. Despite this, our $t_{\rm cool}/t_{\rm visc}$ criterion remains a valid predictor of whether a cold disk condenses out of the accretion flow. A detailed study of multiphase accretion flows with thermal conduction, magnetic fields, and radiative cooling is left for future.
 
\subsection{Cooling Induced State Transitions}

\begin{figure*}
\centering
\includegraphics[scale=0.5]{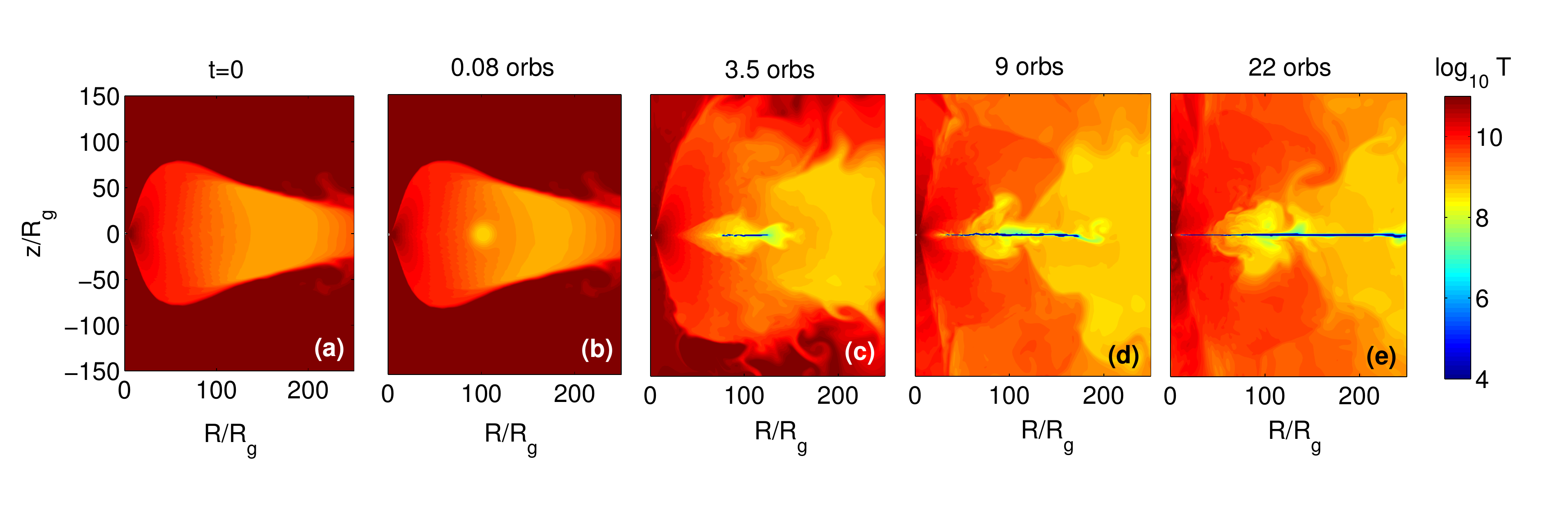}
\caption{Transition from a geometrically thick to a geometrically thin disk induced by addition of mass. The grid resolution is $512^2$.}
\label{fig:brems1e10_thick2thin}
\end{figure*}

\begin{figure*}
\centering
\includegraphics[scale=0.5]{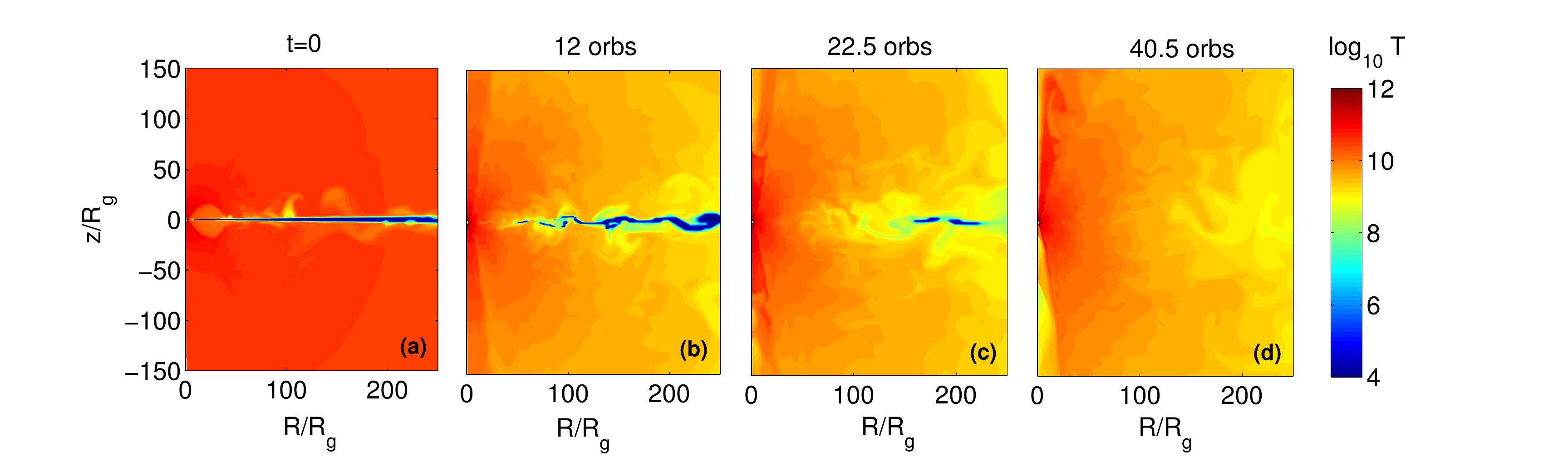}
\caption{Transition from a geometrically thin to a geometrically thick flow due to viscous exhaustion of the thin disk. The grid resolution is $256^2$.}
\label{fig:brems1e10_thin2thick}
\end{figure*}

There are many observational evidences of black hole binary systems that exhibit more than one X-ray spectral state \citep{remi06,fen04}. The spectra from accretion disks surrounding these BHs switch from a hard X-ray dominated state to a disk-dominated soft state, and back to a hard state on a timescale of months to years. Supermassive BHs in galactic centers are also expected to undergo such transients but on much longer timescales because the basic timescales are proportional to the BH mass. The black-body dominated soft state is well-fit by the standard Shakura-Sunyaev thin disk and the hard state can be described by a hot RIAF. The transition from a RIAF to a thin disk state can be induced by adding mass to the accretion flow, thereby increasing the density and decreasing the cooling time of the accretion flow such that the critical ratio $t_{\rm cool}/t_{\rm visc}$ becomes less than unity. The transition from a dense thin disk to a RIAF occurs naturally once sufficient mass in the dense accretion flow is accreted and $t_{\rm cool}/t_{\rm visc}$ becomes $\gtrsim 1$ in the inner regions of the accretion flow.

\subsubsection{RIAF to Thin Disk Transition}

A geometrically thin cold disc can only arise from a hot RIAF if the ratio $t_{\rm cool}/t_{\rm visc} \lesssim 1$. Such a cooling-dominated dense state can be brought about by an increase in the mass accretion rate. This can happen due to the variability of wind from the massive companion (e.g., \citealt{eve98}), due to instabilities close to the Roche lobe  (e.g., \citealt{saw86}), or due to viscous instability of accretion flows (say, due to thermal/viscous instability caused by fast opacity variations with temperature close to $10^4$ K; \citealt{mey81}).  A recent example of the variability of mass accretion rate on large scales in AGNs is the discovery of a dense gas cloud falling in the accretion zone of Sgr $A^*$, the Galactic center BH; this cloud is expected to cause a sudden increase in the mass accretion rate followed by a significant brightening of its X-ray and broadband emission \citep{sgrcloud}. Of course, the amount of mass dumped in this case is too small to cause the formation of a thin black-body disk.

The mass and the density of RIAFs in our simulations decrease with time because of accretion. Therefore, to produce a transition from a RIAF to a thin disk state we need to add mass to the accretion flow. To mimic the increase in the mass accretion rate we include a source function in the density evolution equation (Eq. \ref{eq:cont}) given by
\begin{equation}
\label{eq:den_source}
\dot{\rho} (r,\theta) = 9 \frac{\rho_0}{t_{\rm orb}} \exp \left(-\frac{(r\sin \theta - R_0)^2 + (r\cos \theta)^2}{(10R_g)^2}\right),
\end{equation}
where, $\rho_0$ is the initial density normalization, and $R_0$ is the radius where mass addition is centered (chosen to be the center of the initial dense torus at $100 R_g$), and $t_{\rm orb} = 2 \pi/ \Omega_0$ is the orbital time at $R_0$. The scale over which mass is deposited is $10 R_g$ and the mass source function falls steeply beyond that. The mass is added with the local velocity and internal energy density.

The initial condition of our RIAF to thin disk transition run (Fig. 6(a)) is the end state (at 24 orbits) of our low density run with $\rho_0=10^9 m_p$ g cm$^{-3}$ and $\alpha=0.01$ (similar to Fig. 3(a)). Figure 6(b) shows how the addition of mass looks like at early times --- the mass being added at the local temperature ($T \sim 10^{8-9}$ K) in a circular region of radius $10R_g$ about $R_0$.  Since mass addition is continuous, the density of the flow starts increasing, and after a few orbits the cooling time becomes shorter than the viscous time. This leads to the formation of a cold thin disk having $T \sim 10^4$ K. With time, as the density increases further because of mass addition, there is more efficient cooling of the hot coronal gas into a geometrically thin disk which extends both inwards and outwards. The final steady state shows a geometrically thin disk extending very close to the black hole, surrounded by a hot low density corona with $T\sim 10^9$ K. The expansion of the thin disk towards the black hole takes place at the viscous timescale ($t_{\rm visc}$) at the inner edge of the disk in Figure 6(c). With more mass added, the disk expands viscously both inwards and outwards (Figs. 6(d,e)).

\subsubsection{Thin Disk to RIAF Transition}

The transition from a geometrically thin cold disk to a hot RIAF requires the reduction of mass in the accretion flow such that $t_{\rm cool}/t_{\rm visc} \gtrsim 1$ throughout. This happens naturally if the mass accretion rate in the thin disk at inner radii is larger than the mass addition rate at larger radii. Thus the thin disk to a RIAF transition happens on the viscous timescale at the outer thin disk (mass is dominated by outer radii), which can be quite long. Transient energetic events such as large magnetic flares and explosive nuclear burning are not expected to get rid of sufficient mass from the thin disk because most of the energy is expected to be coupled to the hot corona and not the cold massive disk. 

Figure 7 shows the time evolution of a high density run with a steady thin disk. The initial condition of this run is obtained by evolving our standard $\rho_0=10^{11} m_p$ g cm$^{-3}$ and $\alpha=0.1$ run for 24 orbits. Figure 7 shows the temperature contour plots as transition occurs from a geometrically thin to a geometrically thick accretion flow. This transition happens simply because of exhaustion of mass due to accretion. The thin disk slowly recedes from the inner region until disappearing completely, giving way to a hot geometrically thick, low density corona (RIAF). The viscosity parameter $\alpha$ is chosen to be 0.1 instead of $\alpha=0.01$ in order to accelerate the evolution, which still takes about 40.5 orbits to transform into a RIAF.

\begin{figure*}
\psfrag{p}[][][1.2]{$\nu \propto r^0$}
\psfrag{q}[][][1.2]{$\nu \propto \rho$}
\psfrag{r}[][][1.2]{$\nu \propto r^{1/2}$}
\includegraphics[scale=0.5]{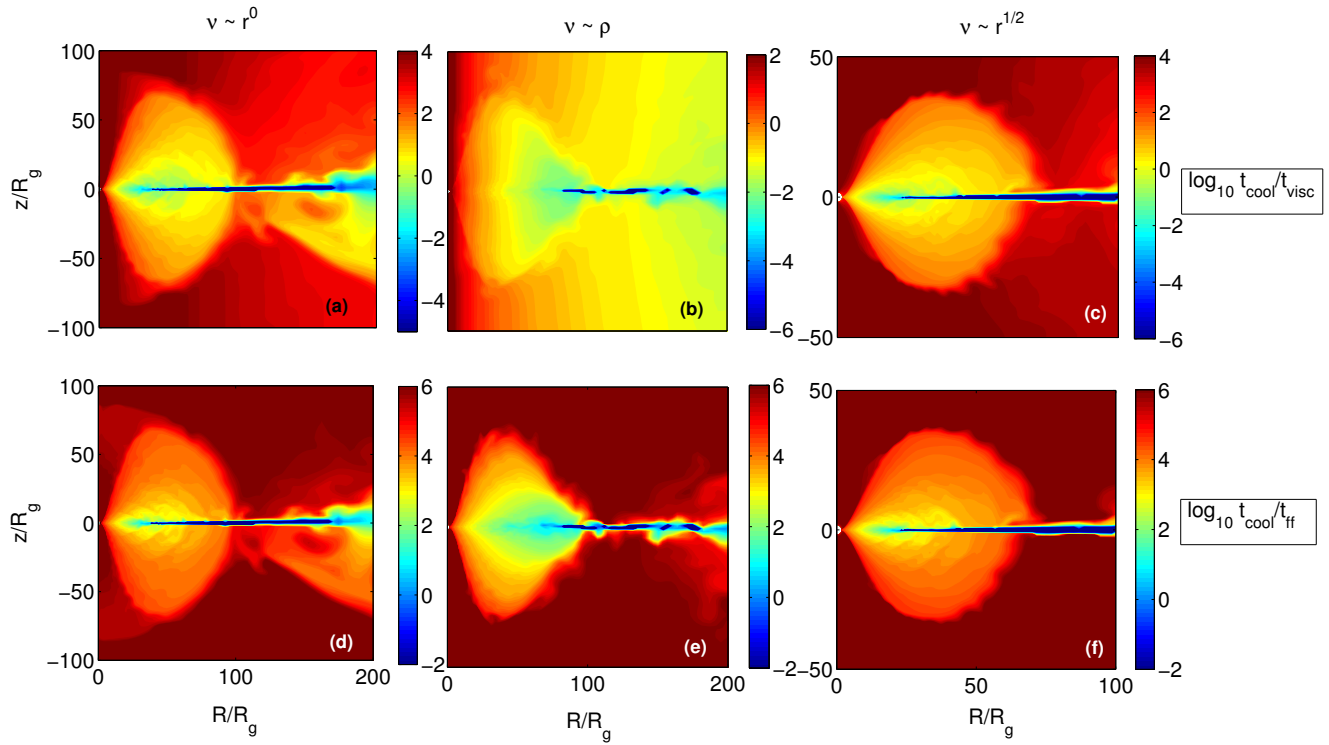}
\caption{(a) (b) (c) Top row - shows the contour plots of $t_{\rm cool}/t_{\rm visc}$ in steady state at late times. (d) (e) (f) Bottom row - shows the plots of $t_{\rm cool}/t_{\rm ff}$ at the same times. From left to right, $\nu=\rm constant$, $\nu \propto \rho$ and $\nu \propto r^{1/2}$. The grid resolution is $512^2$ in all cases. Note that the plots on the rightmost panel are zoomed-in.}
\label{fig:brems1e10_tcooltvisc}
\end{figure*}

\subsection{Density Upper Limit in Corona}
\label{sec:corona}

\begin{figure*}
\psfrag{m}[][][1.5]{$\dot{m}$}
\psfrag{s}[][][1.2]{Mass ($M_\odot$)}
\includegraphics[scale=0.5]{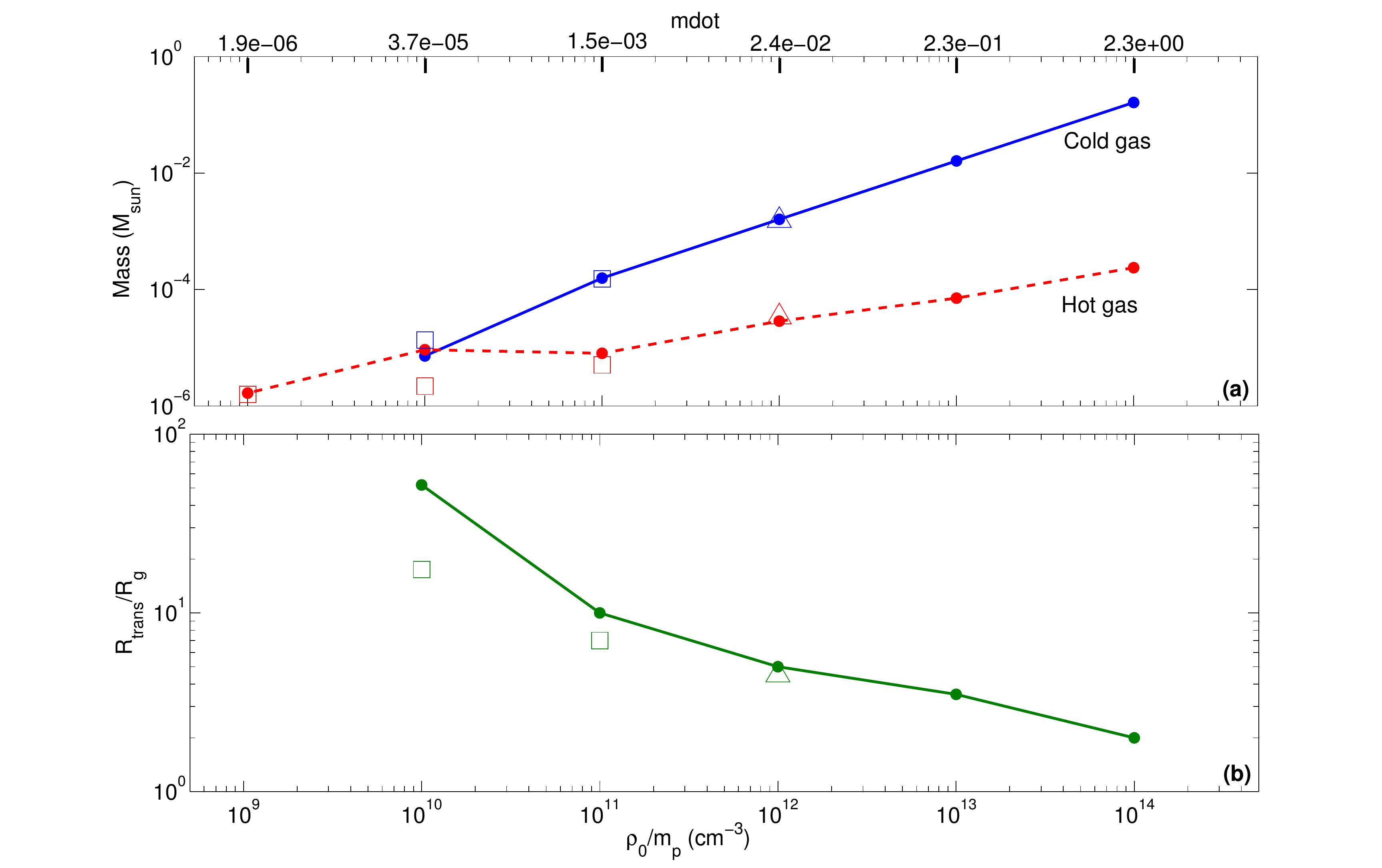}
\caption{(a) Variation of mass in the hot phase ($> 10^5$K; red dashed line) and in the cold phase ($<10^5 $ K; blue solid line) with $\rho_0$. (b) The transition radius $R_{\rm trans}$ as a function of $\rho_0$. The x-axis on the top panel gives the corresponding accretion rate $\dot{m} = \dot{M}/\dot{M}_{\rm Edd}$ through the inner boundary in quasi-steady state, where the Eddington accretion rate is $\dot{M}_{\rm Edd}=6.8 \times 10^{24}~\rm g~s^{-1}$ for a $4\times 10^6 M_{\odot}$ black hole. All runs use $\alpha=0.01$. The filled circles, open triangles and open squares correspond to grid resolutions of $128^2$, $256^2$ and $512^2$ respectively. } 
\label{fig:brems1e10_mass_hbyr}
\end{figure*}

In the previous sections we have established the condition necessary for the formation of a cold, geometrically thin disk, namely, $t_{\rm cool}/t_{\rm visc} \lesssim 1$. We have also observed that if the initial density is high enough to trigger bremsstrahlung cooling, the final steady state consists of both a geometrically thin cold disk and an inner hot low density corona. The next question which arises is about the distribution of matter in the hot and cold phases. More precisely, what is the gas density distribution in the hot corona around a cold thin disk?

In order to address this question we have carried out simulations with different forms of accretion viscosity: $\nu \propto r^{1/2}$, $\nu \propto r^0$, and $\nu \propto \rho$. The magnitude of the viscosity is chosen such that $t_{\rm cool}/t_{\rm visc} \lesssim 1$ in the RIAF and hence a cool disk condenses. A subtle point to note is that this ratio should be calculated in the RIAF state {\em before} a thin disk condenses. The ratio $t_{\rm cool}/t_{\rm visc}$ can be $\ll 1$ in the corona {\em at the radius where a thin disk exists} because here the corona is in thermal equilibrium (e.g., Figure 8(b)). Any imbalance between the cooling and heating of the corona will be compensated by condensation on (or evaporation from) the thin disk. Figure 8 compares the ratio of cooling to viscous time ($t_{\rm cool}/t_{\rm visc}$) and the ratio of cooling to free-fall time ($t_{\rm cool}/t_{\rm ff}$) at late times, for these runs with different forms of viscosity. We observe that the ratio $t_{\rm cool}/t_{\rm visc}$ in the inner hot corona surrounding the cold thin disk is different for the different forms of $\nu$; this is expected since $t_{\rm visc}$ depends on $\nu$. However, the minimum value of $t_{\rm cool}/t_{\rm ff}$ in the corona close to the thin disk, for all the three cases, is quite similar ($\sim 10-100$). Thus we can conclude that the ratio of the cooling time to free-fall time, $t_{\rm cool}/t_{\rm ff} \sim 10-100$, corresponds to the upper limit on the density of the hot corona, irrespective of viscosity. 

The corona is cooling due to bremsstrahlung and is heated due to viscous dissipation (Eq. (3)). Thus it is in rough thermal balance in steady state. Even if the corona is in thermal balance, mass cannot be added to it indefinitely, as a dense corona is expected to undergo multiphase cooling due to local thermal instability; excess mass should condense and fall  onto the cold disk (\citealt{gas13}). This is analogous to what happens in dense cores of galaxy clusters, which, as shown by \cite{icm}, self-adjust such that the ratio of the thermal instability timescale ($t_{\rm TI}$) to the free-fall timescale ($t_{\rm ff}$) satisfies $t_{\rm TI}/t_{\rm ff} \gtrsim 10$ everywhere, in order to achieve global thermal balance. The minimum $t_{\rm cool}/t_{\rm ff}$ ratio in our accretion disk corona is $10-100$ and respects the above limit, irrespective of viscosity.  A detailed investigation of subtleties due to differences in the cooling time and the thermal instability time, spherical versus cartesian gravity (see \citealt{mcc12,icm}), etc.  are beyond the scope of this paper. The heated corona can lead to outflows but we have not evolved our simulations long enough to verify this.

The distribution of gas in the disk vs. the corona is further illustrated by Figure 9(a), which shows the mass of the cold ($T \sim 10^{4-5}$ K) and hot ($ 10^5 \le T \le 10^{10.5}$ K) gas in the accretion flow, as $\rho_0$ is varied. Note that the cold gas represents the geometrically thin disk, while the hot gas makes up the low density, geometrically thick corona. We observe that for $\rho_0 < 10^{10}m_p$ $\rm g~cm^{-3}$ there is no cold gas. However, as $\rho_0$ increases and crosses the $t_{\rm cool}/t_{\rm visc} \lesssim 1$ threshold, the mass of the cold disk increases steeply. On the contrary, the mass of hot gas increases much slowly compared to the mass of the cold gas. As we have already discussed, the coronal density has an upper limit (corresponding to $t_{\rm cool}/t_{\rm ff} \gtrsim 10-100$), so the increase in the hot gas mass with increasing $\rho_0$ occurs because of a bigger size of the corona, and because of a larger fraction of gas at higher densities (and lower temperatures). 

\subsection{Transition Radius}

An important indicator of the structure of the accretion flow is the transition radius ($R_{\rm trans}$), the radius at which the hot accretion flow at smaller radii ($R<R_{\rm trans}$) transitions to a thin disk at larger radii ($R>R_{\rm trans}$). There is so far no accurate method to estimate the exact location of $R_{\rm trans}$, both theoretically and observationally \citep{nar08}. For quantitative estimates we define $R_{\rm trans}$ to be the radius where $H/R \approx 0.02$, recalling that the geometrically thick RIAF has $H/R \approx 0.1$ and $H/R \lesssim 10^{-4}$ for a thin disk. We have already seen that for our simulations with $\rho_0 < 10^{10}m_p$ $\rm g~cm^{-3}$ there is no cold gas and the entire flow is a hot RIAF. 
(If the initial torus is placed much farther out, then a thin disk can form at larger radii even for a smaller initial $\rho_0$.)
Thus, the transition radius for $\rho_0 = 10^{9}m_p$ $\rm g~cm^{-3}$ can be considered to be $> 100 R_g$. As the density of the flow increases (which is equivalent to an increase in the mass accretion rate), we note that $R_{\rm trans}$ decreases monotonically, i.e., the cold thin disk moves closer to the black hole. This is illustrated by Figure 9(b). Thus, a smaller $R_{\rm trans}$ signifies a cooling/thin-disk dominated flow, while a larger $R_{\rm trans}$ indicates that the flow is dominated by a hot radiatively inefficient corona.

\section{A Model for Black Hole Transients}

\begin{figure*}
\psfrag{ha}[cc][][1.1]{$\dot{m}_{\rm hot}=6.6\times 10^{-5}$}
\psfrag{ca}[cc][][1.1]{$\dot{m}_{\rm cold}=0$}
\psfrag{hb}[cc][][1.1]{$\dot{m}_{\rm hot}=-0.06$}
\psfrag{cb}[cc][][1.1]{$\dot{m}_{\rm cold}=0.05$}
\psfrag{hc}[cc][][1.1]{$\dot{m}_{\rm hot}=0.02$}
\psfrag{cc}[cc][][1.1]{$\dot{m}_{\rm cold}=0.09$}
\psfrag{hd}[cc][][1.1]{$\dot{m}_{\rm hot}=8.2\times 10^{-4}$}
\psfrag{cd}[cc][][1.1]{$\dot{m}_{\rm cold}=5.5\times 10^{-3}$}
\includegraphics[scale=0.5]{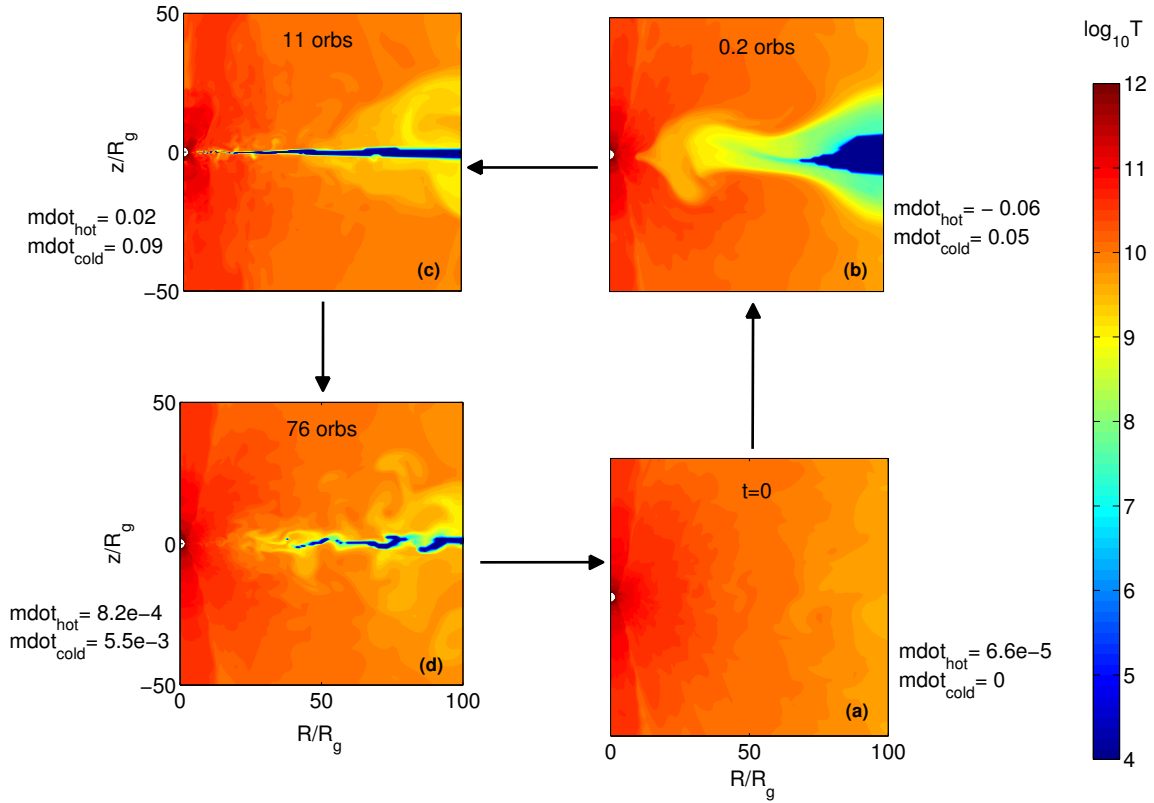}
\caption{Complete state transition diagram. (a) Low-hard state (b) Intermediate hard state (c) High-soft state (d) Intermediate soft state. The intermediate soft state transitions to the hard state at $\gtrsim 100$ orbs. $\dot{m}_{\rm hot}$ and $\dot{m}_{\rm cold}$ are the accretion rates in units of $\dot{M}_{\rm Edd}$ for the hot and cold gas respectively. The grid resolution is $256^2$.}
\label{fig:brems1e10_statetr}
\end{figure*}

In previous sections we established the condition for the formation of a thin disk and showed how cooling and viscous evolution brings about state transitions. In this section we show  the results of a single simulation with $\alpha=0.1$ and initial density $\rho_0 = 10^{9}m_p$ $\rm g~cm^{-3}$ (Run A.1n9), which demonstrates the complete transition from a RIAF to a cold thin disk and back to a RIAF. Previous works, like ours, have identified the low-hard and high-soft states with a RIAF and a geometrically thin black body disk, respectively, but have not studied the time evolution of BH transients in detail (e.g., \citealt{esi97}).

We initialize a hot RIAF with $T \sim 10^{10}$ K everywhere (Figure 10(a)). This initial condition is obtained by evolving a $\rho_0=10^9 m_p$ g cm$^{-3}$ torus initial condition for several orbits. The cooling time is long in this case due to the low density, and hence there is no signature of a thin disk. Next, we add mass (after about 0.02 orbits at $R_0$) to the flow according to the density source term in Eq. \ref{eq:den_source}. Figure 10(b) shows the onset of cold gas formation due to the increase in density and the decrease in cooling time brought about by mass addition. As the mass accumulates (much) faster than the accretion time, the density of the flow keeps increasing and eventually a thin cold disk (Figure 10(c)) is formed which extends very close to the black hole ($\lesssim 10 R_g$). The condensation of the initial thin disk occurs on the fast cooling timescale. The initial thin disk spreads inward on the  viscous time at the inner edge of the thin disk in Figure 10(b). 

Once a steady thin disk is formed (after about 11 orbits), we stop the mass addition and the disk is simply allowed to evolve viscously. As accretion continues, sufficient mass is exhausted from the disk and the thin disk recedes far away from the central black hole (Figure 10(d)). As more and more mass is accreted into the black hole, the density of the flow drops, causing the cooling time to become longer than the viscous time, eventually resulting in the formation of a RIAF (in about 100 orbits from the start). The luminosity in this RIAF state continues decreasing on the viscous timescale at the circularization radius ($\sim R_0$ in the present case). Thus we return to the original state with which we started (similar to Figure 10(a)). Note that the thin disk state is the most long-lived of the transient states, since it takes a significant amount of time to remove sufficient mass from the disk through accretion, as this process takes place at a viscous time at the outer edge of the steady thin disk, which is pretty long. Recall that we are using $\nu \propto r^{1/2}$ as the functional form for the viscosity, for both the thin disk and the RIAF. Whereas, in reality, the viscous diffusion coefficient for a thin disk is expected to be $(H/R)^2 \sim 10^{-4}$ times smaller than in a RIAF. Thus, the black-body state is expected to be even longer lived than what is indicated in Figure 10.

X-Ray astronomers popularly use the Hardness-Intensity diagram (HID) to describe the overall evolution of the spectral states of black hole XRBs, which are observationally known to be transient sources \citep{remi06,fender12}. Typically in the HID, a black hole transient source transitions between various spectral states classified according to their X-ray luminosity and X-ray hardness tracing a `q'-shape, as shown, for example, by the 2002/2003 outburst of GX 339-4 \citep{belloni05}. One can draw a one-to-one correspondence between the various spectral states and the ones shown in Figure 10 as follows --- low-hard state (Figure 10(a)), hard intermediate state (Figure 10(b)), high-soft state (Figure 10(c)) and soft intermediate state (Figure 10(d)). This can be done, since Figure 10 also gives an estimate of the mass accretion rate (an indicator of the luminosity) corresponding to each state. We must mention that the observed transients show short-timescale oscillations in hardness during the intermediate state which are probably not caused by variations in the mass accretion rate but are likely due to dynamical timescale magnetic processes in the corona (e.g., reconnection). Our simulations lack magnetic fields and therefore cannot directly address the phenomenology of radio emission, quasi-periodic oscillations (QPOs), and fast transients observed in the intermediate state.

We have already shown that the condition for the flow to be a RIAF is $t_{\rm cool}/t_{\rm visc} \approx 3nk_BT\nu/n^2\Lambda(T)r^2 \gtrsim 1$. This condition can be recast in terms of the accretion rate using the following scalings. From the equation of continuity one can write the number density $n \approx \dot{M}/2\pi \nu m_p H$. The gas temperature for a RIAF is nearly virial and hence $k_BT \sim GMm_p/r$. Also, we write $\dot{M}$ in terms of the Eddington accretion rate $\dot{M}_{\rm Edd}=4\pi GMm_p/\eta \sigma_T c$, where $\sigma_T$ is the Thomson scattering cross section and we choose the radiative efficiency parameter $\eta$ to have the standard value 0.1 \citep{nar95}.
Thus, the condition for the accretion flow to be a RIAF at $10R_g$, putting in all the numbers, is $\dot{m} \lesssim 0.1~ \alpha^2$, where $\dot{m}=\dot{M}/\dot{M}_{\rm Edd}$ (this was first pointed out by \citealt{ree82}). This indicates the existence of a critical accretion rate above which cooling will start dominating and a thin disk will form.

The simulation in Figure 10 corresponds to $\alpha=0.1$. Thus, the condition for the flow to be a RIAF becomes $\dot{m} \lesssim 10^{-3}$, which is indeed the case in Figure 10. In all the panels $\dot{m}_{\rm cold}$ represents the accretion rate of the cold gas, while $\dot{m}_{\rm hot}$ represents the accretion rate of the hot gas. The `accretion' rates are calculated by taking the difference of mass in the cold and hot phases at two consecutive snapshots. Figure 10(a) shows no cold gas, and $\dot{m}_{\rm hot}=6.6\times 10^{-5} \ll 10^{-3}$; this confirms that this flow is a low-luminosity RIAF with a low accretion rate (corresponding to the quiescent state). The state shown in Figure 10(b) corresponds to the hard intermediate state since mass in the hot gas becomes quite substantial, cold gas has just started to condense. The magnitude of the mass `accretion' rate is $\approx 0.001$ (this is actually the mass addition rate due to the source  term in Eq. \ref{eq:den_source}), of order the critical value for the formation of a thin disk ($10^{-3}$). The hot phase accretion rate $\dot{m}_{\rm hot}$ is negative since the rate at which mass is added to the hot phase is smaller than the rate at which hot gas condenses into the cold phase. Figure 10(c) corresponds to a steady thin disk state with $\dot{m}_{\rm cold}=0.09$, which is greater than $\dot{m}_{\rm hot}=0.02$; thus it is a cooling dominated flow. Also, $\dot{m} \approx 0.1 \gg 10^{-3}$ indicates that this state has a higher luminosity and a prominent thin disk, namely it is the high-soft state. Figure 10(d) corresponds to the soft intermediate state with $\dot{m}_{\rm cold}=5.5\times 10^{-3}$, which is still greater than $\dot{m}_{\rm hot}=8.2\times 10^{-4}$. While the thin disk has receded significantly in this state, there are still some signatures of the thin disk.

Note that, although in our simulations we mainly study the geometrical structure of the accretion flow, we can still get an idea about the luminosity and the degree of hardness/softness of emitted X-rays. The soft photons are radiated by a geometrically thin, optically thick multi-color blackbody disk. The hot RIAF, on the other hand, is associated with hard X-rays, and the corresponding spectrum is dominated by a power-law; the power law arises because of the extension of the corona (and its temperature) over several decades in radius. Note that the conversion of the mass accretion rate ($\dot{m}$) into a luminosity is not straightforward. The thin disk luminosity is $\sim GM\dot{M}/2R_{\rm in}$ where $R_{\rm in}$ is the inner radius of the cold disk; it not only depends on $\dot{M}$ but also on the inner truncation radius of the thin disk. Similarly, the emissivity of the RIAF is $n^2 \Lambda(T)$, so the RIAF luminosity is proportional to $\dot{m}^2$ and subdominant compared to the energy released via accretion $\sim GM\dot{M}/2R_{\rm LSO}$ (where $R_{\rm LSO}$ is the last stable orbit); most of the RIAF accretion energy goes into outflows (e.g., Fig. 1 in \citealt{chu05}). 

From the above considerations we note that Figure 10 describes a complete state transition, (a) $\rightarrow$ (b) $\rightarrow$ (c) $\rightarrow$ (d) $\rightarrow$ (a), which can be explained by variability in the mass accretion rate and viscous evolution of the accretion flow. Thus, very interestingly, our model for state transitions qualitatively reproduces the features of the HID. The observed timescales in lightcurves and hardness evolution can be used to constrain the circularization radius of the transient thin disk, its outer radius, the transition radius, and the total mass added in the accretion event.

\section{Conclusions}
The key conclusions of the paper are:
\begin{itemize}

\item The condition for the transition of a RIAF to a cold, geometrically thin disk is that the cooling time be shorter than the viscous time, i.e., 
$t_{\rm cool}/t_{\rm visc} \lesssim 1$ (see Fig. 3). When the density of the accretion flow exceeds the critical density corresponding to $t_{\rm cool}/t_{\rm visc} \lesssim1$, cooling processes such as bremsstrahlung are triggered, resulting in a globally stable two-phase steady state consisting of an inner hot RIAF and an outer geometrically thin disk. The accretion flow with cooling is sensitive to the Shakura-Sunyaev viscosity parameter $\alpha$. A higher $\alpha$, which implies a shorter $t_{\rm visc}$, requires a higher density to trigger cooling and hence to form a thin disk (see Fig. 4). 

\item While whether a cool disk condenses is determined by $t_{\rm cool}/t_{\rm visc}$, the ratio of the cooling time and the free-fall time determines the structure of the disk corona.
The minimum ratio of the cooling time to the free-fall time in the corona close to the thin disk is found to be $t_{\rm cool}/t_{\rm ff} \sim 10-100$ (see Fig. 8). This ratio is independent of the kinematic viscosity $\nu$ and indicates an upper limit for the density of the hot coronal phase, which is in thermal balance. If mass is added to the corona such that this  density threshold is exceeded, the excess mass simply condenses and falls onto the thin disk. In the thin disk state the accretion flow has a subdominant but a substantial corona, which should produce some hard radiation even in the high-soft state (see Fig. 9(a)).

\item With the increase of the mass accretion rate, the transition radius $R_{\rm trans}$ between the inner hot RIAF and the outer thin disk decreases monotonically, i.e., the thin disk moves closer to the black hole with an increasing mass accretion rate (see Fig. 9(b)). 

\item We have qualitatively simulated transitions between various X-ray spectral states of black hole transients and their relative positions in the Hardness-Intensity diagram/q-plot (see Fig. 10). Our model gives an estimate of the overall timescales in XRB transients. From our model we conclude that the transition from a hot RIAF to a geometrically thin disk occurs due to an increase in the mass accretion rate. This transition progresses on the viscous timescale at the inner edge $R_{\rm in}$ of the initially formed thin disk (which can be quite fast if the circularization radius is small). The timescale of transition from a low-hard to a high-soft  state can thus be estimated to be $t_{\rm visc} \approx R_{\rm in}^2/\alpha c_sH$. For a $10M_{\odot}$ BH with the inner disk at $R_{\rm in} \sim 10^3 R_g$, $\alpha=0.1$, the sound speed of $10^5$ K cold disk $c_s \sim 3\times 10^6 {\rm cm~ s^{-1}}$ and $H/R_{\rm in} \sim 0.01$, we estimate  $t_{\rm visc, R_{in}} \sim {\rm 10 ~days}$. Similarly, our model shows that the transition from a cold, thin disk to a geometrically thick, hot RIAF occurs at the viscous timescale at the outer edge $R_{\rm out}$ of the steady state thin disk (which is much longer). For the same XRB considered above, if $R_{\rm out}$ is $\sim 10^4 R_g$, we obtain $t_{\rm visc, R_{out}} \sim {\rm 1 ~year}$ (here we have assumed that temperature $T \propto r^{-3/4}$). The increase in luminosity in the hard state ((a)$\rightarrow$(b) in Fig. 10) should happen on the mass addition timescale, which can be pretty short (of the order of dynamical time). These timescales match pretty well with generic observations of BH XRBs. Since Roche-lobe filling XRBs have larger circularization radii (and hence longer viscous times) compared to the XRBs accreting massive stellar winds, the global timescales in the q-plot for stellar wind accreting BH XRBs should be shorter.

\end{itemize}

\section*{Acknowledgments}

UD thanks CSIR, India for financial support. The numerical simulations were carried out on the computer cluster supported by the start-up grant of PS at IISc. We thank Banibrata Mukhopadhyay for encouragement and helpful discussions.

\label{lastpage}

\end{document}